# Imaging With Nature: Compressive Imaging Using a Multiply Scattering Medium

Antoine Liutkus[1], David Martina[1,2], Sébastien Popoff[1], Gilles Chardon[3], Ori Katz[1,2], Geoffroy Lerosey[1], Sylvain Gigan[1,2], Laurent Daudet[1], Igor Carron[4]

[1]Institut Langevin, ESPCI ParisTech, Paris Diderot Univ., UPMC Univ. Paris 6, CNRS UMR 7587, Paris, France
[2]Laboratoire Kastler-Brossel, UMR8552 CNRS, Ecole Normale Supérieure, Univ. Paris 6, Collège de France, 24 rue Lhomond, 75005 PARIS
[3]Acoustics Research Institute, Austrian Academy of Sciences, Vienna.
[4]TEES SERC, Texas A&M University.

**Abstract**
The recent theory of compressive sensing leverages upon the structure of signals to acquire them with much fewer measurements than was previously thought necessary, and certainly well below the traditional Nyquist-Shannon sampling rate. However, most implementations developed to take advantage of this framework revolve around controlling the measurements with carefully engineered material or acquisition sequences. Instead, we use the natural randomness of wave propagation through multiply scattering media as an optimal and instantaneous compressive imaging mechanism. Waves reflected from an object are detected after propagation through a well-characterized complex medium. Each local measurement thus contains global information about the object, yielding a purely analog compressive sensing method. We experimentally demonstrate the effectiveness of the proposed approach for optical imaging by using a 300-micrometer thick layer of white paint as the compressive imaging device. Scattering media are thus promising candidates for designing efficient and compact compressive imagers.

**Introduction**

Acquiring digital representations of physical objects - in other words, *sampling* them - was, for the last half of the 20th century, mostly governed by the Shannon-Nyquist theorem. In this framework, depicted in Fig. 1(a), a signal is acquired by N regularly-spaced samples whose sampling rate is equal to at least twice its bandwidth. However, this line of thought is thoroughly pessimistic since most signals and objects of interest are not only of limited bandwidth but also generally possess some additional *structure* (15). For instance, images of natural scenes are composed of smooth surfaces and/or textures, separated by sharp edges.

Recently, new mathematical results have emerged in the field of Compressive Sensing (or Compressed Sensing, CS in short) that introduce a paradigm shift in signal acquisition. It was indeed demonstrated by Donoho, Candès, Tao and Romberg (5,10,2) that this additional structure could actually be exploited *directly at the acquisition stage* so as to provide a drastic reduction in the number of measurements without loss of reconstruction fidelity.

For CS to be efficient, the sampling must fulfill specific technical conditions that are hard to translate into practical design guidelines. In this respect, the most interesting argument featured very early on in (5,10,2) is that a *randomized* sensing mechanism yields perfect reconstruction with high probability. As a matter of convenience, hardware designers have created physical systems that *emulate* this property. This way, each measurement gathers information from all parts of the object, in a controlled but pseudo-random fashion. Once this is achieved, CS theory provides good reconstruction guarantees.

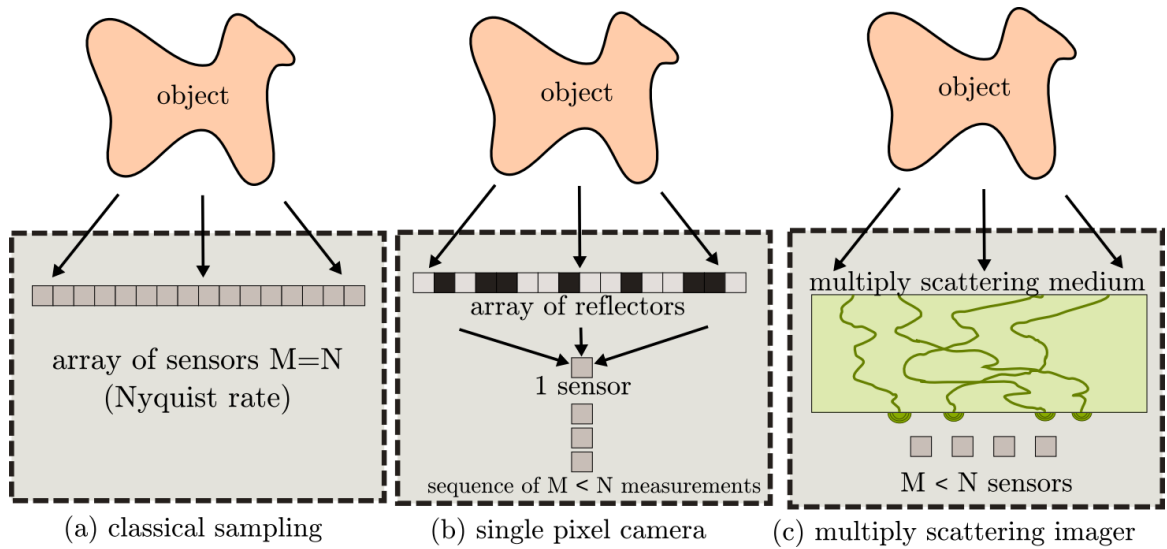

**Fig. 1 Concept**. (a) Classical Nyquist-Shannon sampling, where the waves originating from the object, of size N, are captured by a dense array of M=N sensors. (b) The "Single Pixel Camera" concept, where the object is sampled by M successive random projections onto a single sensor using a digital multiplexer. (c) Imaging with a multiply scattering medium. The M sensors capture, in a parallel fashion, several random projections of the original object. In cases (b) and (c), sparse objects can be acquired with a low sensor density M/N<1.

In the past few years, several hardware implementations capable of performing such random compressive sampling were introduced (6,14,18,23,24,25,27,34,45). In optics, these include the single pixel camera (14), which is depicted in Fig. 1(b), and uses a digital array of micromirrors (abbreviated DMD) to sequentially reflect different random portions of the object onto a single photodetector. Other approaches include phase modulation with a spatial light modulator (25), or a rotating optical diffuser (45). The idea of random multiplexing for imaging has also been considered in other domains of wave propagation. CS holds much promise in areas where detectors are rather complicated and expensive such as the THz or far infrared. In this regards, there have been proposals to implement CS imaging procedures in the THz using random pre-fabricated masks (6), DMD or SLM photo-generated contrast masks on semi-conductors slabs (35) and efforts are also pursued on tunable metamaterial reflectors (7). Recently, a carefully engineered metamaterial aperture was used to generate complex RF beams at different frequencies (23).

However, these CS implementations come with some limitations. First, these devices include carefully engineered hardware designed to achieve randomization, via a DMD (14), a metamaterial (23) or a coded aperture (27). Second, the acquisition time of most implementations can be large because they require the *sequential* generation of a large number of random patterns.

In this work, we replace such man-made *emulated* randomization by a natural multiply scattering material, as depicted in Fig. 1(c). Whereas scattering is usually seen as a time-varying nuisance, for instance when imaging through turbid media (40), the recent results of wave control in stable complex material have largely demonstrated that it could also be exploited, for example so as to build focusing systems that beat their coherent counterparts in terms of resolution (41,42). Such complex and stable materials are readily available in several frequency

ranges - they were even coined in as one-way physical functions for hardware cryptography (30). In the context of CS, such materials perform an efficient randomized multiplexing of the object into several sensors and hence appear as *analog* randomizers. The approach is applicable in a broad wavelength range and in many domains of wave propagation where scattering media are available. As such, this study is close in spirit to earlier approaches such as the random reference structure (9), the random lens imager (18), the metamaterial imager (23), or the CS filters proposed in (28) for microwave imaging. They all abandoned digitally controlled multiplexors as randomizers. Still, we go further in this direction and even drop the need for a designer to *craft* the randomizer.

Compressive sampling with multiply scattering material has several advantages. First, it has recently been shown that they have an optimal multiplexing power for coherent waves (19), which consequently makes them optimal sensors within the CS paradigm. Second, these materials are often readily available and require very few engineering. In the domain of optics for example, we demonstrate one successful implementation using a 300μm layer of Zinc Oxide (ZnO), which is essentially white paint. Third, contrarily to most aforementioned approaches, this sensing method provides the somewhat unique ability to take a scalable number of measurements in parallel, thus strongly reducing acquisition time. In practice, if about, say, 500 samples are required to reconstruct a given image using CS principles, our proposed approach permits to acquire them simultaneously, whereas state-of-the-art systems such as the single pixel camera require a sequence of 500 individual measurements.

On practical grounds, the use of a multiply scattering material in CS raises several ideas that we consider in this study. First, the random multiplexing achieved through multiple scattering must be measured *a posteriori*, since it is no longer known *a priori* as in engineered random sensing. This calibration problem has been the topic of recent studies in the context of CS (22) and we propose here a simple least squares calibration procedure that extends our previous work (31,32). Second, the use of such a measured Transmission Matrix (TM) induces an inherent uncertainty in the sensing mechanism, that can be modeled as noise in the observations. As we show both through extensive simulations and actual experiments, this uncertainty is largely compensated by the use of adequate reconstruction techniques. In effect, the imager we propose almost matches the performance of idealized sub-Nyquist random sensing.

**Theoretical background**

In its simplest form, CS may be understood as a way to solve an underdetermined linear inverse problem. Let $x$ be the object to image, understood as a $N \times 1$ vector, and let us suppose that $x$ is only observed through its multiplication $y$ by a known measurement matrix $H$, of dimension $M \times N$, we have $y = Hx$. Each one of the $M$ entries of $y$ is thus the scalar product of the object with the corresponding row of $H$. When there are fewer measurements than the size of the object, i.e. $M < N$, it is impossible to recover $x$ perfectly without further assumptions, since the problem has infinitely many solutions. However, if $x$ is known to be *sparse*, meaning that only a few of its coefficients are nonzero (such as stars in astronomical images), and provided $H$ is sufficiently random, $x$ can still be recovered uniquely through sparse optimization techniques (15).

In a signal processing framework, the notion of *structure* may also be embodied as sparsity in a known representation (15). For example, most natural images are not sparse, yet often yield near-sparse representations in the wavelet domain. If the object $x$ is known to have some sparse or near-sparse representation $s$ in a known basis $B$ ($x = Bs$), then it may again be possible to recover it from a few samples, by solving $y = HBs$, provided $H$ and $B$ obey some technical conditions such as *incoherence* (2,5,10,15,16).

In practice, when trying to implement Compressive Sensing in a hardware device, fulfilling this *incoherence* requirement is nontrivial. It requires a way to deterministically *scramble* the information somewhere between the object and the sensors. Theory shows that an efficient way to do this is by using *random measurement matrices $H$ or $HB$* (2,5,10). Using such matrices, it can indeed be shown (16) that the number of samples required to recover the object is mostly governed by its sparsity $k$, i.e. the number of its nonzero coefficients in the given basis. If the coefficients of the $M \times N$ measurement matrix are independent and identically distributed (i.i.d.) with respect to a Gaussian distribution, perfect reconstruction can be achieved with only $O\left(k \log(N/k)\right)$ measurements. Furthermore, many algorithms are available, for instance Orthogonal Matching Pursuit (OMP) or Lasso (15,36), which can efficiently perform such reconstruction under sparsity constraints.

## Using natural complex media as random sensing devices

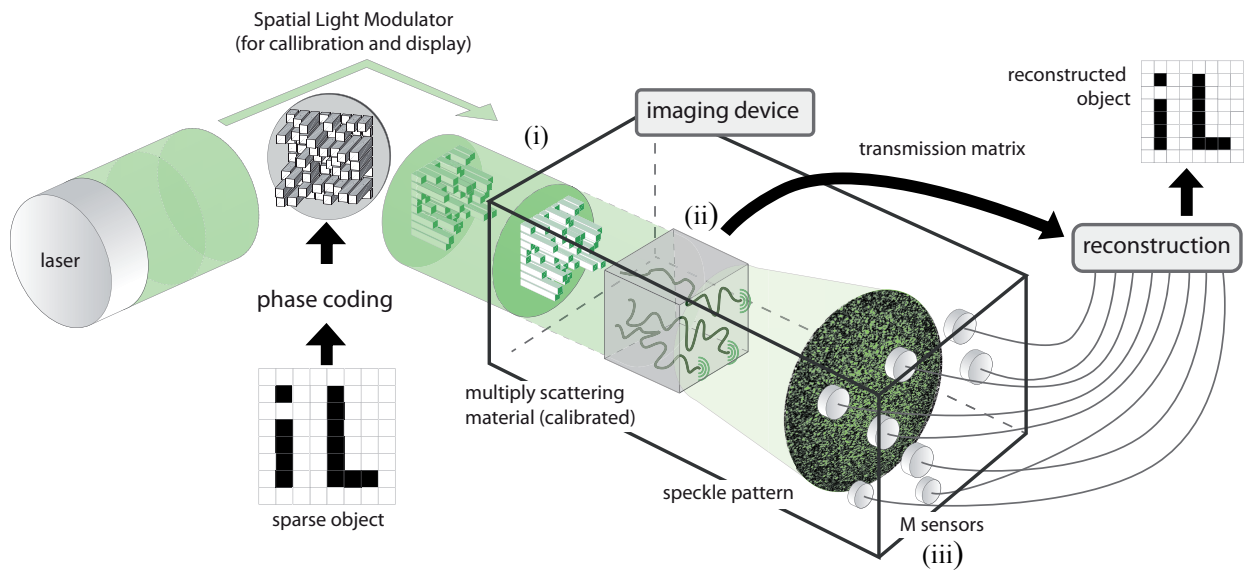

**Fig. 2 Experimental setup** for compressive imaging using multiply scattering medium. Within the imaging device, waves coming from the object (i) go through a scattering material (ii) that efficiently multiplexes the information to all $M$ sensors (iii). Provided the transmission matrix of the material has been estimated beforehand, reconstruction can be performed using only a limited number of sensors, potentially much lower than without the scattering material. In our optical scenario, the light coming from the object is displayed using a spatial light modulator.

Our approach is summarized in Fig. 1(c) and its implementation in an optical experiment is depicted in Fig.2. The coherent waves originating from the object and entering the imaging system propagate through a multiply scattering medium. Within the imager, propagation produces a seemingly random and wavelength-dependent interference pattern called speckle on the sensors plane. The speckle figure is the result of the random phase variations imposed on the waves by the propagation within the multiply scattering sample (20). Scattering, although the realization of a random process, is deterministic: for a given input, and as long as the medium is stable, the interference speckle figure is fully determined and remains constant. In essence, the complex medium acts as a highly efficient analog multiplexer for light, with an input-output response characterized by its transmission-matrix (31,32). We highlight the fact that the multiple scattering material is not understood here as a nuisance occurring between the object and the sensors, but rather as a desirable component of the imaging system itself. After propagation, sensing takes place using a limited number $M < N$ of sensors.

Let $x$ and $y$ denote the $N \times 1$ and $M \times 1$ vectors gathering the value of the complex optical field at discrete positions before and after, respectively, the scattering material. It was confirmed experimentally (31,32) that any particular output $y_m$ can be efficiently modeled as a linear function of the $N$ complex values $x_n$ of the input optical field:

$$y_m = \sum_{n=1}^{N} h_{mn} x_n,$$

where the mixing factor $h_{mn} \in \mathbb{C}$ corresponds to the overall contribution of the input field $x_n$ into the output field $y_m$. All these factors can be gathered into a complex matrix $[H]_{mn} = h_{mn}$ called the *Transmission Matrix* (TM), which characterizes the action of the scattering material on the propagating waves between input and output. The medium hence produces a very complex but deterministic mixing of the input to the output, that can be understood as spatial multiplexing. This linear model, in the ideal noiseless case, can be written more concisely as:

$$y = Hx.$$

As can be seen, each of the $M$ measurements of the output complex field may hence be considered as a scalar product between the input and the corresponding row of the TM. If multiply scattering materials have already been considered for the purpose of *focusing*, thus serving as perfect "opaque lenses" (41,42), the main idea of the present study is to exploit them for compressive imaging. In other wavelength domains than optics, analogous configurations may be designed to achieve CS through multiple scattering. For instance, a collection of randomly packed metallic scatterers could be used as a multiply scattering media from the microwave domain up to the far infrared, and the method proposed here could allow imaging at these frequencies with only a few sensors. A similar approach could be used to lower the number of sensors in 3D ultrasound imaging using CS through multiple scattering media.

In our optical experimental setup, we used a Spatial Light Modulator (an array of $N = 1024$ micromirrors, abbreviated as SLM) to calibrate the system and also to display various objects, using a monochromatic continuous wave laser as light source.

During a first *calibration* phase, which lasts a few minutes and needs to be performed only once, a series of controlled inputs $x$ are emitted and the corresponding outputs $y$ are measured. The TM can be estimated through a simple least-squares error procedure, which generalizes the method proposed in (31,32), as detailed in the supplementary material below. In short, this calibration procedure benefits from an arbitrarily high number of measurements for calibration, which permits to better estimate the TM. It is important to see here that the need for this calibration step is the main practical inconvenient of the proposed approach compared to more classical CS imagers based on pseudo-random projections. Indeed, those latter do control the TM perfectly whereas it is only estimated in our case. However, in many cases, this calibration step only involves a very standard least-squared error estimation of the linear mapping between the input and output of the scattering material (31,32), which is done in less than 1 minute in our optical experimental setup, to be compared with the time required to design a pseudo-random projection machinery. Still, it is understood that to apply the proposed methodology at other wavelengths, one needs a way to estimate the linear mapping between the input and the output of the scattering medium. If it is easily done in optics, it may not be so straightforward at other wavelengths, e.g. when only the intensity of the output is available.

After calibration, the scattering medium can be used to perform CS, using this estimated TM as a measurement matrix[1]. As demonstrated in our results section, using such an estimated TM instead of a perfectly controlled one does yield very good results all the same, while bringing important advantages such as ease of implementation and acquisition speed. Hence, even if the proposed methodology does require the introduction of a supplementary calibration step, this step comes at the cost of a few mandatory supplementary calibration measurements rather than at the cost of performance. This claim is further developed in our results and methods sections.

For a TM to be efficient in a CS setup, it has to correctly scramble the information from all of its inputs to each of its outputs. It is known that a matrix with i.i.d Gaussian entries is an excellent candidate for CS (12) and the TM of optical multiple scattering materials were recently shown to be well approximated by such matrices (19). The rationale for this fact is that the transmission of light through an opaque lens leads to a very large number of independent scattering events. Even if the total transmission matrix that links the whole input field to the transmitted field shows some non-trivial mesoscopic correlations (1), recent studies proved that these correlations vanish when controlling/measuring only a random partition of input/output channels (19). In our experimental setup, the number of sensors is very small compared to the total number of output speckle grains and we can hence safely disregard any mesoscopic correlation.

Several previous studies (31,32) have shown on experimental grounds that TMs were close to i.i.d. Gaussian by considering their spectral behavior, i.e. the distribution of their eigenvalues. As a consistency check, we also verified that our experimentally-obtained TMs are close to Gaussian i.i.d., through a complementary study of their *coherence,* which is the maximal correlation between their columns with values between 0 and 1. Among all the features that were proposed to characterize a matrix as a good candidate for CS (3,8,44), coherence plays a special role because it is easily computed and because a low coherence is sufficient for good recovery performance in CS applications (11,38,39,43), even if it is not necessary (4). In Fig. 3(a), we display one actual TM obtained in our experiments. In Fig. 3(b), we compare its coherence with the one of randomly generated i.i.d. Gaussian matrices. The similar behavior confirms the results and discussions given in (19,31), but also suggests that TMs are good candidates in a CS setup, as will be demonstrated below.

---

[1] In our experiment, the same SLM used for calibration is then used as a display to generate the sparse objects. This approach is not restrictive as any sparse optical field or other device capable of modulating light could equivalently be used at this stage.

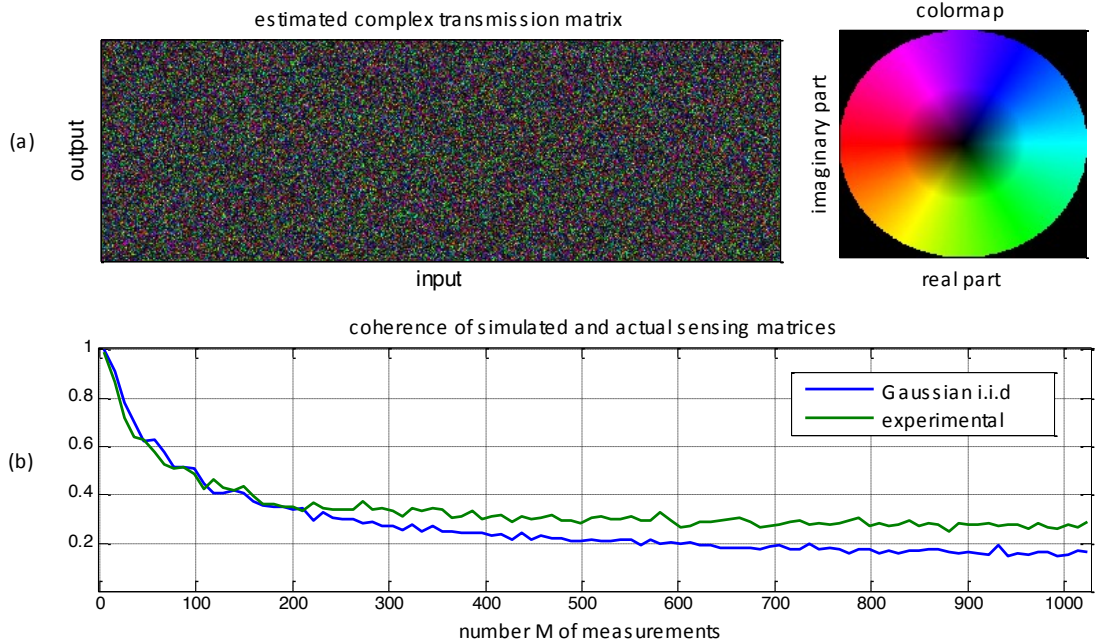

**Fig. 3 Experimentally measured Transmission Matrix** (TM). (a) TM for a multiply scattering material as obtained in our experimental study. (b) Coherence of sensing matrices as a function of their number $M$ of rows, for both a randomly generated Gaussian i.i.d. matrix, and an actual experimental TM. Coherence gives the maximal colinearity between the columns of a matrix. The lower, the better is the matrix for CS.

**Results and discussion**

During our experiments, we measured the reconstruction performance of the imaging system, when the image to reconstruct is composed of $N = 32 \times 32 = 1024$ pixels, using a varying number $M$ of measurements. In practice, we use a CCD array, out of which we select $M$ pixels. These are chosen at random in the array, with an exclusion distance equal to the coherence length of the speckle, in order to ensure uncorrelated measurements. Details of the experiments can be found in the methods section below. For each sparsity level $k$ between 1 and $N$, a sparse object[2] with only $k$ nonzero coefficients was displayed under $P = 3$ different random phase illuminations. The corresponding outputs were then measured and fed into a Multiple Measurement Vector (MMV) sparse recovery algorithm (9). For each sparsity level, 32 such independent experiments were performed.

Reconstruction of the sparse objects was then achieved numerically using the $M \times P$ measurements only. The TM used for reconstruction is the one estimated in the calibration phase. In order to demonstrate the efficiency and the simplicity of the proposed system, we used the simple Multichannel Orthogonal Matching Pursuit algorithm (21) for MMV reconstruction. It should be noted that more specialized algorithms may lead to better performance and should be considered in the future.

---

[2] Since our SLM can only do phase modulation, we used a simple trick as in (33) to simulate actual amplitude objects, based on two phase-modulated measurements. See the supplementary material on this point. Those virtual measurements may be replaced by the use of an amplitude light modulator and are anyways replaced by the actual object to image in a real use-case.

Examples of actual reconstructions performed by our analog compressive sampler are shown on Fig. 4. As can be seen, near-perfect reconstruction of complex sparse patterns occur with sensor density ratios $M/N$ that are much smaller than in classical Shannon-Nyquist sampling ($M = N$). An important feature of the approach is its universality: reconstruction is also efficient for objects that are sparse in the Fourier domain.

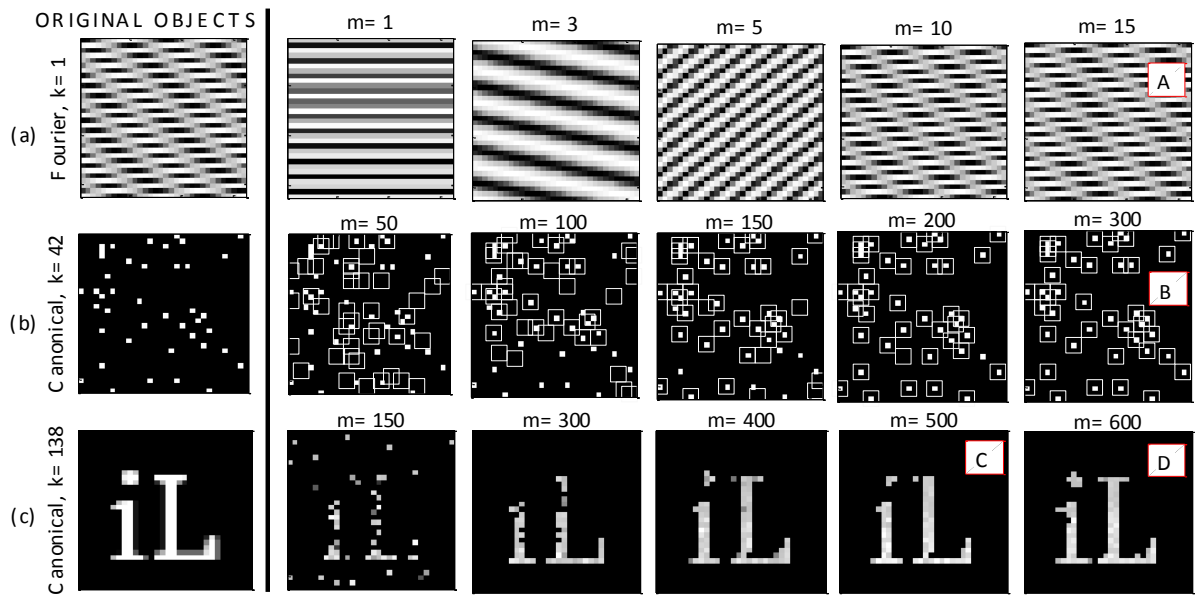

**Fig. 4 Imaging results.** Examples of signals, which are sparse either in the Fourier or canonical domain (left), along with their actual experimental reconstruction using a varying number of measurements. (a) Fourier-sparse object (b-c) canonical sparse objects. In (b), small squares are the original object and large squares are the reconstruction. In all cases, the original object contains 1024 pixels and is thus sampled with a number $M$ of sensors much smaller than $N$. A, B, C and D images are correspondingly represented in the phase transition diagram of Fig. 5.

The performance of the proposed compressive sampler for all sampling and sparsity rates of interest is summarized on Fig. 5, which is the main result of this paper. It gives the probability of successful reconstruction displayed as a function of the sensor density $M/N$ and relative sparsity $k/M$. Each point of this surface is the average reconstruction performance for real measurements over approximately 50 independent trials. As can be seen, this *experimental* diagram exhibits a clear "phase transition" from complete failure to systematic success. This thorough experimental study largely confirms that the proposed methodology for sampling using scattering media indeed reaches very competitive sampling rates that are far below the Shannon-Nyquist traditional scheme.

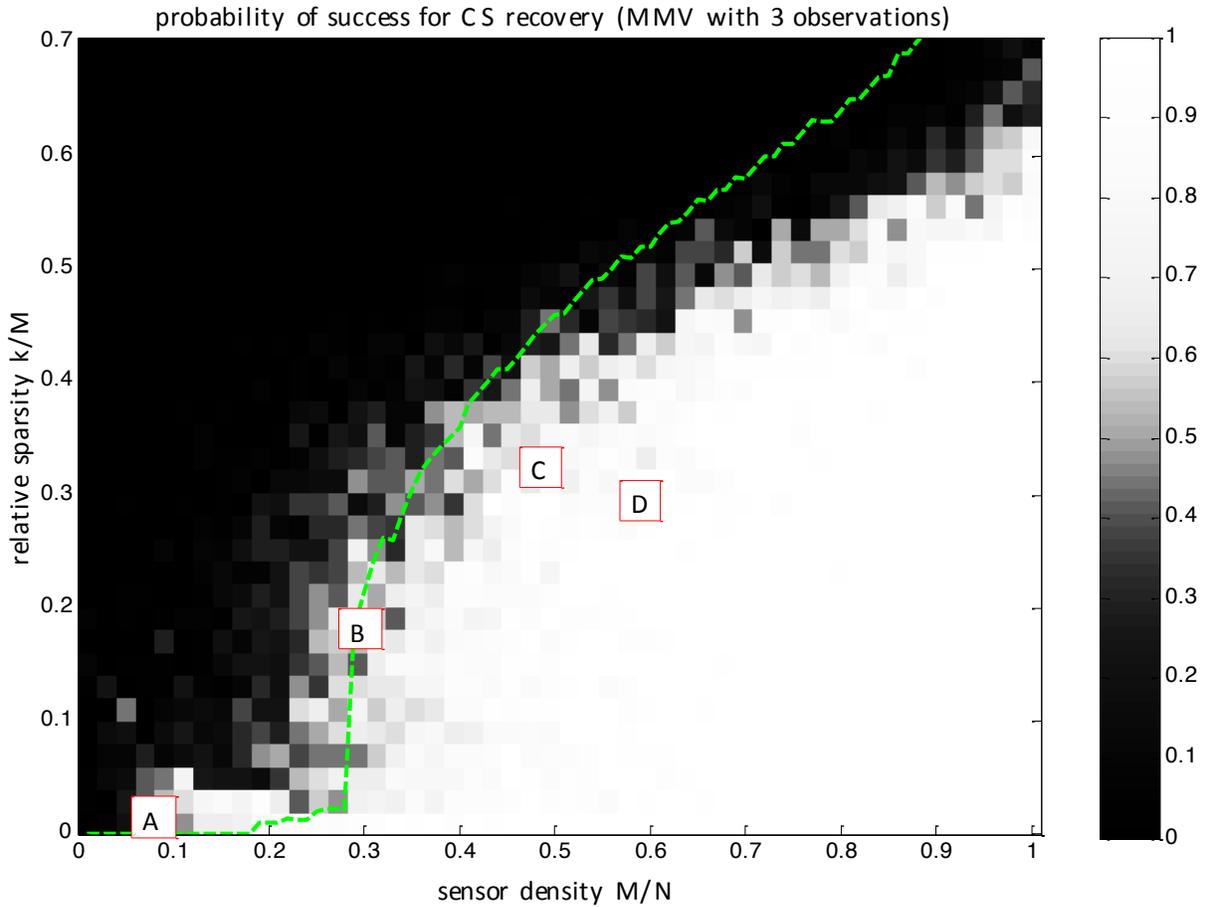

**Fig. 5 Probability of success for CS recovery.** Experimental probability of successful recovery (between 0 and 1) for a k-sparse image of $N$ pixels via $M$ measurements. On the x-axis is displayed the sensor density ratio $M/N$. A ratio of 1 corresponds to the Nyquist rate, meaning that all correct reconstructions found in this figure beat traditional sampling. On the y-axis is displayed the relative sparsity ratio $k/M$. A clear phase transition between failure and success is observable, which is close to that obtained by simulations (dashed line), where exactly the same experimental protocol was conducted with simulated noisy observations both for calibration and imaging. Boxes A, B, C and D locate the corresponding examples of Fig. 4. Each point in this 50x50 grid is the average performance over approximately 50 independent measurements. This figure hence summarizes the results of more than $10^5$ actual physical experiments.

The phase transition observed on Fig. 5 appears to be slightly different from the ones described in the literature (12,13). The main reason for this fact is that this diagram concerns reconstruction under $P = 3$ Multiple Measurement Vectors (MMV) instead of the classical Single Measurement Vector (SMV) case. This choice, which proves important in practice, is motivated by the fact that MMV is much more robust to noise than SMV (17). In order to compare our experimental performance to its numerical counterpart, we performed a numerical experiment whose 50% success-rate transition curve is represented by the dashed green line. The transmission matrix is taken as i.i.d Gaussian. The measurement matrix is estimated with the same calibration procedure as in the physical experiment. Each measurement, during calibration and imaging, is contaminated by additive Gaussian noise of variance 3%. Performance obtained in this idealized situation is close to that obtained in our practical setup, for this level of additive noise.

**Conclusion**

In this study, we have demonstrated that a simple natural layer of multiply scattering material can be used to successfully perform compressive sensing. The compressive imager relies on scattering theory to optimally dispatch information from the object to all measurement sensors, shifting the complexity of devising CS hardware from the design, fabrication and electronic control to a simple calibration procedure.

As in any hardware implementation of CS, experimental noise is an important issue limiting the performance, especially since it impacts the measurement matrix. Using baseline sparse reconstruction algorithms along with standard least-squares calibration techniques, we demonstrated that successful reconstruction exhibits a clear phase transition between failure and success even at very competitive sampling rates. The proposed methodology can be considered to be a truly analog compressive sampler and as such, benefits from both theoretical elegance and ease of implementation.

The imaging system we introduced has many advantageous features. First, it enables the implementation of an extremely flat imaging device with few detectors. Second, this imaging methodology can be implemented in practice with very few conventional lenses in the setup (26). This is a strong point for implementation in domains outside optics where it is hard to fabricate lenses. Indeed, the concept presented here can directly be used in other domains of optics such as holography, but also in other disciplines such as THz, RF or ultrasound imaging. Third, similarly to recent work on metamaterials apertures, non-resonant scattering materials work over a wide frequency range and have a strongly frequency-dependent response. Fourth, unlike most current compressive sensing hardware, this system gives access to many compressive measurements in a parallel fashion, drastically speeding up acquisition. These advantages come at the simple cost of a calibration step, which amounts to estimate the Transmission Matrix of the scattering material considered. As we demonstrated, this can be achieved by simple input/output mapping techniques such as linear least-squares and needs to be done only once.

While conventional direct imaging can be thought as an embarassingly parallel process that does not exploit the structure of the scene, in contrast most current CS hardware (such as the single pixel camera) require a heavily sequential process that does take into account the structure of the scene. Our approach borrows from the best of both acquisition processes, in that it is both embarassingly parallel and takes into account the structure of the scene.

**Acknowledgments:** This work was supported by the European Research Council (Grant N°278025), the Emergence(s) program from the City of Paris, and LABEX WIFI (Laboratory of Excellence within the French Program "Investments for the Future") under references ANR-10-LABX-24 and ANR-10-IDEX-0001-02 PSL*. G.C. is supported by the Austrian Science Fund (FWF) START-project FLAME (Y 551-N13). O.K. is supported by the Marie Curie intra-European fellowship for career development (IEF) and the Rothschild fellowship. I.C. would like to thank the Physics arXiv Blog for drawing his attention to opaque lenses.

# Supplementary material

## Experimental setup

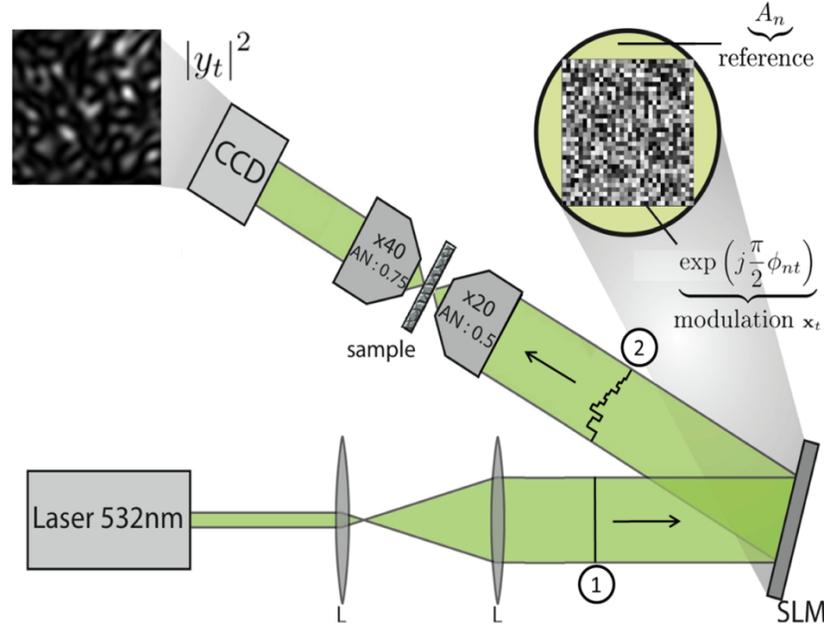

**Fig. S1** Experimental setup for light diffusion through a scattering medium (31). The coherent wavefront from a laser is modulated using a SLM and transmitted through the medium. A CCD camera measures the wavefront at the output of the medium

For one particular measurement $t$, the experiment is displayed on figure S1. A laser beam is enlarged by a couple of (L-L) lenses and the resulting wavefront (1 on figure S1) is partly reflected by a Spatial Light Modulator (SLM) and partly by its support, respectively leading to the modulated wavefront $x_t$ and the reference wavefront $A$, which is constant over $t$. The SLM is composed of a 32 x 32 matrix of $N = 1024$ segmented micromirrors (Kilo-DM, Boston Micromachines Corporation). Each $n$ of those mirrors is a 300μm square and locally controls the phase $\phi_{nt}$ of $x_{nt} = \exp(\pi \phi_{nt}/2)$. Both $x_t$ and $A$ combine to form the reflected wave (2 on figure S1), whose value at position $n$ is thus $A_n x_{nt}$. This wave is then propagated through an opaque 300μm-thin layer of compacted zinc oxide (ZnO) powder. In this medium, light is reflected many times and finally collected and imaged by a CCD camera. The interested reader is referred to (31,32) for more details concerning the experimental setup. In all the remaining of this study, when we mention a specific number $M$ of measurements, we experimentally refer to a subset of the CCD pixels. Since those pixels are always randomly chosen, the corresponding measurements are equivalent to those performed by $M$ arbitrarily located sensors.

The complex wave $y_{mt}$ at each of the $m = 1\ldots M$ output positions is estimated using 4 intensity measurements done by the CCD camera thanks to the phase-stepping technique, which is detailed in (31,37). Note that in all the following, each complex measurement is hence understood as a combination of 4 intensity measurements. In further studies, *compressive phase*

*retrieval* techniques (29) may be used to directly process intensity measurements $|y|$ instead of $y$ and hence further reduce the acquisition time. For now, we will simply consider the complex output $y$, which are related to $Ax_t$ by:

$$y_{mt} = \sum_{n=1}^{N} \bar{H}_{mn} A_n x_{nt},$$

where $\bar{H}$ is the intrinsic TM of the scattering media. Since the reference wave $A$ is kept fixed during the whole process thanks to the stability of the laser beam, it can safely be merged with $\bar{H}$ so as to yield

$$H_{mn} = \bar{H}_{mn} A_n,$$

which will subsequently be called the TM, although it depends both on the scattering medium and the laser input. The slight abuse of notation is largely justified by the very high stability of lasers available on the market. Its estimation rather than $\bar{H}$ is sufficient to proceed to compressed sensing using scattering media. Note however that since $A_n$ stays constant over different measures $t$, it is possible through calibration to identify $\bar{H}$ from $H$ (31,32). In any case, the complex output $y$ is given by:

$$y_{mt} = \sum_{n=1}^{N} H_{mn} x_{nt},$$

which can be written in a more compact matrix-form as $Y = HX$, where $Y = [y_{mt}]_{m,t}$ and $X = [x_{nt}]_{n,t}$ are $M \times T$ and $N \times T$ matrices, respectively, whereas $H$ is the (complex) $M \times N$ TM.

**Estimation of the Transmission Matrix**

In (31,32), Popoff et al. propose to estimate the TM $H$ using an orthonormal basis as input and hence having $X$ as a $N \times N$ matrix. The choice of the Hadamard basis to this purpose is judicious since all its entries are $\pm 1$, which leads to $x_{nt} = \pm j$. Therefore, if $B_N$ denotes the $N \times N$ Hadamard basis, the measured matrix $Y$ is $Y = HB_N$. If $I_N$ is defined as the $N \times N$ identity matrix, one of the properties of $B_N$ is to be its own inverse, leading to $B_N B_N = I_N$ and hence $H = YB_N$. This very simple procedure leads to a straightforward estimation of the TM $H$.

However, a better estimation of the TM is possible, provided more calibration measures are done, i.e. by choosing $T > N$. In that case, $Y = HX$ still holds but $X$ is not an orthonormal basis nor its own inverse. However, $H$ can still be estimated straightforwardly through Least-Squares as:

$$\hat{H} = YX^H \left( XX^H \right)^{-1}$$

where $\cdot^H$ denotes Hermitian (conjugate) transpose. This formula is actually a special case of a much more general setting, where noisy observations are accounted for and where estimation of the TM is performed through Least-Squares estimation.

In any case, in our experimental setup, instead of using a single $N \times N$ Hadamard matrix as $X$, the input matrix $X$ for calibration was built as $X = \exp(j\Phi)$, with $\Phi$ being the horizontal concatenation of $B_N$ and a large random $N \times 5N$ matrix with independent entries, uniformly distributed on the interval $[0; 2\pi]$. Then, after measurements have been performed, $H$ is estimated through the formula above.

A clear limitation of the approach is that estimation of the TM requires the linear outputs $Y = HX$. Even if those linear output may be obtained using phase-stepping techniques in several wavelengths, there are scenarios where only their magnitude may be available. More sophisticated techniques may be used in that case to estimate the TM using such measurements, with good performance in practice. Such approaches are the topic of current work.

**Virtually sparse intensity inputs**

In this section, we describe how the input data to the proposed imaging system was generated. S*parse* signals are zero most of the time and only scarcely nonzero. However, due to the particular experimental setup, where light is modulated using a phase-only SLM, we cannot consider signals, which are sparse in the Dirac (canonical) domain. Indeed, this would amount to having $x_{nt} = 0$ most of the time, which is impossible because all $x_{nt}$ have the same amplitude: our SLM performs *phase* and not *amplitude* modulation.

However, we can use a simple trick that was already considered by Popoff et al. in (33) to generate an arbitrary (virtual) phase and amplitude object from a phase modulator. We use the same technique to build *virtual sparse objects* that are constructed as follows. First, build a $N \times 1$ random phase vector $\phi_r$, called *reference*, and measure the corresponding $M \times 1$ complex output $y_r$. Second, randomly choose $k$ entries in $\phi_r$, called the *support* and set their values as new random phases to build the vector $\phi_s$, which is identical to $\phi_r$ except for only $k$ entries. The corresponding complex output $y_r$ is measured, and thanks to the linearity of the optical propagation, the difference $y = y_r - y_s$ corresponds to the complex output of the system for the *sparse virtual input object* $x = x_r - x_s$. Using this procedure, we were able to measure the output of the system for sparse input vectors of arbitrary sparsity $k$.

We highlight the fact that this way to build sparse inputs is required only because we used a SLM to control the input wavefront and not because of intrinsic limitations of the imaging method we propose. On the contrary, we emphasize that such virtual measurements actually lead to additional (doubled) noise, making the imaging process only more difficult.

Additionally and as done in (33), we were able to measure $P = 3$ several outputs corresponding to different *illuminations* of the same virtual sparse object. This was achieved by using the same support for $P$ different reference phases. In essence, we thus settle in the Multiple Measurement Vector paradigm, abbreviated as MMV (9) and depicted in figure S2. The total number of complex measurements for each trial is hence $3M$, used to estimate $3N$ values of the input field.

In our experiments, we repeated this procedure so as to build a very large number of virtual objects of varying sparsity, from $k = 1$ to $k = N$, along with their corresponding outputs.

**Algorithm for reconstruction using compressed sensing**

Suppose for now that the considered input wave fronts $x$ are sparse in the canonical domain, thus being virtual objects in our experimental setup as described above. We suppose that their sparsity $k$ is known and that the complex outputs $Y$ of the system for $P$ different illuminations of the same object are available, as depicted on figure S2.

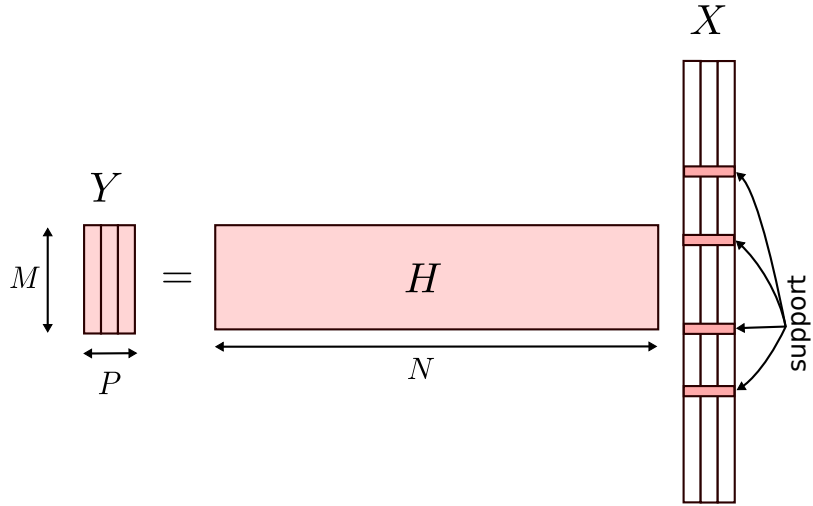

**Fig. S2** The $P$ measurements vectors of $Y$ correspond to the output of the system to $P$ different sparse inputs having the same support. For inputs that are sparse in a base $B$ different from the canonical one, $H$ is simply replaced by $HB$.

We measure the $M \times P$ matrix $Y$, corresponding to the concatenation of $P$ outputs and given by $Y = HX$, where $X$ is the unknown input matrix to estimate and $H$ is the TM. Provided $H$ has been estimated through calibration, any algorithm capable of estimating a sparse vector given $M$ random projections of it can be used for the purpose of estimating $X$. In our experiments, we made use of basic multichannel Orthogonal Matching Pursuit, OMP (21), which is straightforward to implement. Even if more sophisticated methods may be used, we purposefully focused on the most simple and widely accepted approach to CS, since the purpose of this study was not to concentrate on alternative estimation methods, which may rather be the topic of future work. Note that when $P = 1$, the whole procedure simply becomes equivalent to classical OMP.

Once the input $X$ has been estimated, the correlation of its support with ground truth is computed and estimation is said to be successful if this correlation lies above 0.9, meaning that at least 90% of the original support has been identified. When we are considering vectors that are not sparse in the Dirac (canonical) basis but in an alternative basis $B$, notably the Fourier basis as explained above, the same procedure can be applied using $HB$ instead of $H$ as a measurement matrix.

We applied this procedure for approximately 25000 different inputs, corresponding to a large range of sparsity $k$ from 1 to $N = 1024$, and for many different values for the number $M$ of measurements, so as to yield a complete phase transition as found by Donoho and Tanner (12,13), displayed on Fig. 5. Each cell of this figure gives the average observed performance for the corresponding set of $(k, M)$ parameters over approximately 50 independent trials.

An identical experiment was then performed with measurements that are obtained by simply multiplying the sparse inputs by a synthesized i.i.d. Gaussian matrix and further adding a noise whose average amplitude is set to 17% of the observed average amplitude of the synthesized clean output. This matrix is estimated and used for CS in exactly the same manner as for the experimental data, in effect comparing performance of the presented imager with that of an idealized random sensor whose matrix would be unknown but estimated using noisy data. The transition curve for this idealized case is displayed in figure 4 as a dashed line.

**Fourier-sparse inputs**

Even if virtual objects are a good way to simulate objects, which are sparse in the canonical domain with arbitrary sparsity $k$, it is desirable to test the proposed imaging system using direct measurements of sparse objects. To this purpose, we measured the output of the system when the input $x$, of constant modulus, is sparse in the 2D-Fourier domain. In other words, it is easy to build $x$ as a 2D planewave so that its modulus is constant while only one element of its Fourier transform is non-zero and corresponds to its wave number.

Although this procedure is simple, it is difficult to generalize it for arbitrary sparsity $k$, since it is not straightforward to build 2D wavefronts of constant modulus and arbitrary sparsity in the Fourier domain. Given one sparsity level (either $k = 1$ or $k = 20$), performance of the imaging method is evaluated as a function of the number $M$ of measurements and the results are displayed on Fig. S3.

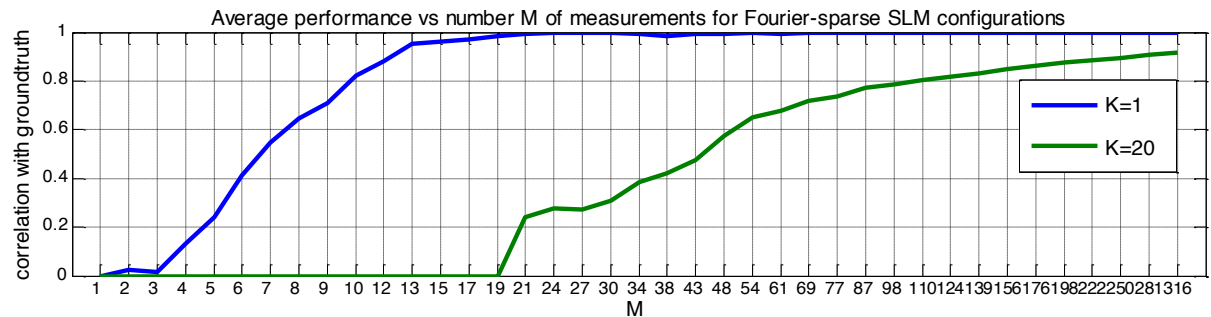

**Fig. S3** Compressed sensing performance for the recovery of signals which are sparse in the Fourier domain. The two curves correspond to the probability of success as a function of the number of measurements $M$, when the unknown signal is either a plane wave ($k = 1$) or the superposition of many planewaves ($k = 20$). Each point is the average of 128 independent trials.

As can be seen on this figure, 15 measurements are sufficient to properly recover the input wavefront of the system, provided it is a planewave. This result demonstrates that the proposed imaging system is indeed universal and that its performance well matches results predicted by CS theory.